\documentclass[twocolumn,showpacs,prb]{revtex4}

\usepackage{graphicx}
\usepackage{dcolumn}
\usepackage{bm}
\usepackage{amssymb}

\begin{document}

\preprint{???}

\title{Aging and scaling laws in $\beta$-hydroquinone-clathrate}

\author{Maikel~C.~Rheinst\"adter}
\altaffiliation[Present address: ]{Institut Laue-Langevin, 6 rue
Jules Horowitz, BP 156, 38042 Grenoble Cedex 9, France}
\email{rheinstaedter@ill.fr}

\affiliation{Technische Physik, Universit\"at des Saarlandes, PSF
1551150, 66041 Saarbr\"ucken, Germany}


\author{Heiko~Rieger}
\affiliation{Theoretische Physik, Universit\"at des Saarlandes,
PSF 1551150, 66041 Saarbr\"ucken, Germany}

\author{Klaus~Knorr}%
\affiliation{Technische Physik, Universit\"at des Saarlandes, PSF
1551150, 66041 Saarbr\"ucken, Germany}

\date{\today}

\begin{abstract}
The dielectric permittivity of the orientational glass
methanol(x=0.73)-$\beta$-hydroquinone-clathrate has been studied
as function of temperature and waiting time using different
temperature-time-protocols. We study aging, rejuvenation and
memory effects in the glassy phase and discuss similarities and
differences to aging in spin-glasses. We argue that the diluted
methanol-clathrate, although conceptually close to its magnetic
pendants, takes an intermediate character between a true
spin-glass and a pure random field system.
\end{abstract}

\maketitle

\section{Introduction}
The long time evolution of the response functions of glassy
systems has attracted a lot of attention and gave and gives rise
to many
discussions\cite{Vincent:1998,Bouchaud:1997,Bouchaud:2000}. The
basic observation is that the behavior of the physical response
functions, like magnetization, polarization or the corresponding
susceptibilities, is time and waiting time dependent. It depends
on a waiting time $t_{\omega}$ spent before a perturbation (a
magnetic or electric field) has been applied or after it has been
switched off. A spin-glass can even memorize some of the features
of the way the system had been prepared. While in the beginning
theoretical and experimental work on {\em physical aging} mainly
concentrated on aging phenomena in polymer glasses
\cite{Struik:1978}, a lot of effort has been made to carry out and
understand experiments in other glassy systems, like
spin-glasses\cite{Jonason:1998}, pure and disordered
ferromagnets\cite{Vincent:2000}, supercooled organic liquids such
as glycerol\cite{Leheny:1998}, relaxor
ferroelectrics\cite{Colla:2000} and (crystalline) dipolar
glasses\cite{Alberici:1998,Alberici:2000}. The differences and the
things in common between aging in these systems are still
intensively discussed. It is not clear which ingredients on a
microscopic scale have to be included to observe aging,
rejuvenation- and memory-effects. In the present paper we present
aging, rejuvenation and memory experiments in a dipolar glass,
namely methanol(x=0.73)-$\beta$-hydroquinone-clathrate.

The quinol HO-C$_6$H$_4$-OH molecules form a hydrogen bonded
R$\bar{3}$ lattice with almost spherical cavities of 4.2 \AA\
diameter. The host lattice tolerates methanol concentrations down
to x=0.35 without collapsing, x being the ratio of filled to
available cavities. Whereas the higher concentrated
methanol-clathrates undergo a first order phase transition into an
antiferroelectrically ordered low-temperature phase, the
clathrates with lower concentrations (x$\leq$0.76) freeze into
dipolar glass states\cite{Woll:2000,Woll:2001}. In
methanol(x=0.73)-$\beta$-hydroquinone-clathrate, the freezing of
the dipoles is at least to some part a collective process as the
molecules do not only see the crystal field of the cavity but also
a contribution from the electric interaction between the
dipoles\cite{Woll:2000,Woll:2001}. In this sense the
methanol(x=0.73)-clathrate is conceptually close to its magnetic
pendants, the spin-glasses. Within the hexagonal basal plane, the
dipoles ("pseudospins") are arranged on a triangular lattice.
Together with the quasi-antiferroelectric long-ranged
dipole-dipole interaction in the basal plane, the dilution leads
to strong frustration and to a large number of metastable states.
Although there is no clear transition into a glassy phase, we find
slow glassy dynamics at low temperatures.

Our first investigation\cite{Kityk:2002} of aging phenomena in
methanol-hydrochinone-clathrates produced qualitative evidence for
aging, rejuvenation and memory effects. In the present article we
investigate the frequency dependence of isothermal aging (Sect.
\ref{isothermal} and \ref{scaling}) and study rejuvenation and
aging in a more quantitative way by following special
time-temperature protocols in Sect. \ref{rejuvenation}.

\section{Experimental results\label{exerimental results}}
Single crystals of methanol(x=0.73)-$\beta$-hydroquinone-clathrate
have been grown from a saturated solution of quinol, methanol and
n-propanol at 313 K. The propanol molecules are not incorporated
into the lattice but merely control the percentage of void
cavities. The samples were prepared in the form of thin parallel
plates ($d\sim 0.5$ mm) with faces perpendicular to the hexagonal
$c$-axis. With gold electrodes deposited on these faces they
formed capacitors with a capacitance of about 0.4 pF. For the
measurement of the frequency-dependent permittivity, a Solartron
impedance analyzer FRA 1260 in combination with an interface
(Chelsea) has been employed. The samples were placed into a
closed-cycle refrigerator with a temperature stability of some mK.

\begin{figure}
\centering
\resizebox{1.00\columnwidth}{!}{\rotatebox{0}{\includegraphics{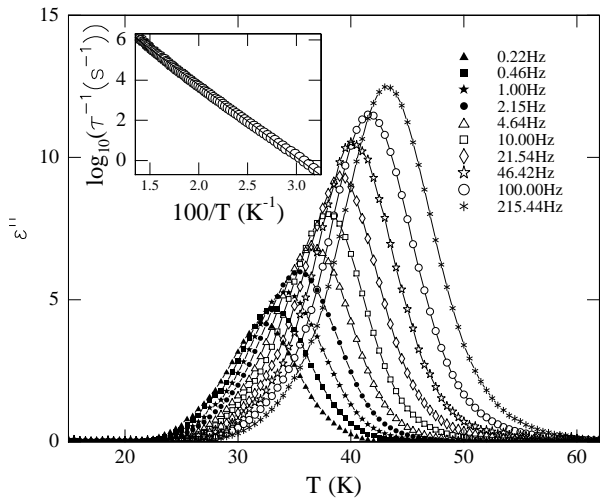}}}
\caption[]{Temperature dependence of the imaginary part of the
dielectric permittivity, $\varepsilon''_c(\nu)$. (From
[\onlinecite{Kityk:2002}]). The inset shows the inverse average
relaxation time $\tau^{-1}$ as a function of 1/T (determined from
the temperature of maximum $\varepsilon_c''$).}
\label{paperbild.eps}
\end{figure}
\begin{figure}
\resizebox{1.00\columnwidth}{!}{\rotatebox{0}{\includegraphics{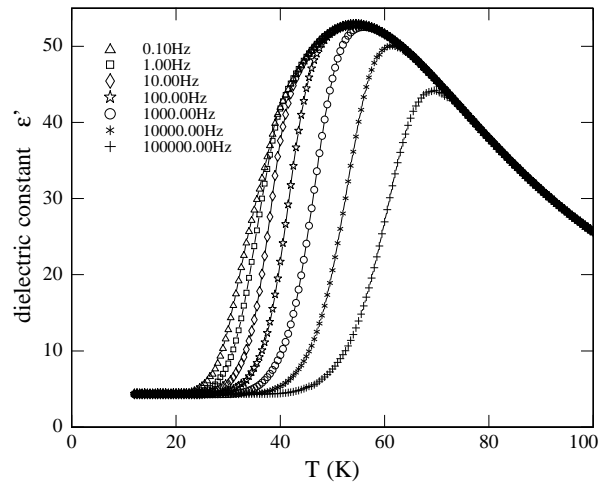}}}
\caption[]{Temperature dependence of the real part of the
dielectric constant, $\varepsilon'_c(\nu)$.}
\label{dielectricconstant.eps}
\end{figure}

We measured real and imaginary part of the dielectric constant,
$\varepsilon_c'(\nu)$ and $\varepsilon_c''(\nu)$, along the
hexagonal c-axis for frequencies from $\nu=0.2$ Hz up to $\nu=200$
Hz. Figure \ref{paperbild.eps} shows the temperature dependence of
the imaginary part, $\varepsilon_c''(\nu)$. The relaxation
processes are characterized by a rather broad distribution of
relaxation times $\tau$ and obey an Arrhenius law $\tau^{-1}\sim
\tau_0^{-1}\exp(-E_A/T)$ (with the "pseudospin" flip-rate
$\tau_0^{-1}=9\cdot 10^{10} s^{-1}$ and an energy barrier
E$_A=828$ K) rather than a Vogel-Fulcher-behavior, see inset of
Fig.\@ \ref{paperbild.eps}. Although there is no clear transition
into a dipolar glass state we find glassy dynamics at low
temperatures. The effective energy barrier E$_A$ decomposes into a
contribution of the crystal field (E$_{loc}$) and a contribution
of the interaction between the dipoles,
E$_A\approx$E$_{loc}+4J_c$, $J_c$ being a coupling
parameter\cite{Woll:2000}. In our sample, the freezing is
dominated by random interactions (E$_{loc}\approx$ 210 K,
4$J_c\approx$ 620 K) and should be close to spin-glass-like.
Figure \ref{dielectricconstant.eps} shows the real part of the
dielectric constant $\varepsilon'_c(\nu)$ for selected frequencies
$\nu$ from $\nu$=0.1 Hz to $\nu$=1 MHz. As can be seen from this
figure it is only below about 40 K that the 0.1 Hz and the 1 Hz
curves split. Thus the 0.1 Hz curve represents the static
dielectric response for temperatures above 40 K. The maximum at
about 55 K is a property of the static permittivity that signals
the onset of short range antiferroelectric ordering.

\subsection{Isothermal aging\label{isothermal}}
\begin{figure}
\centering
\resizebox{1.00\columnwidth}{!}{\rotatebox{0}{\includegraphics{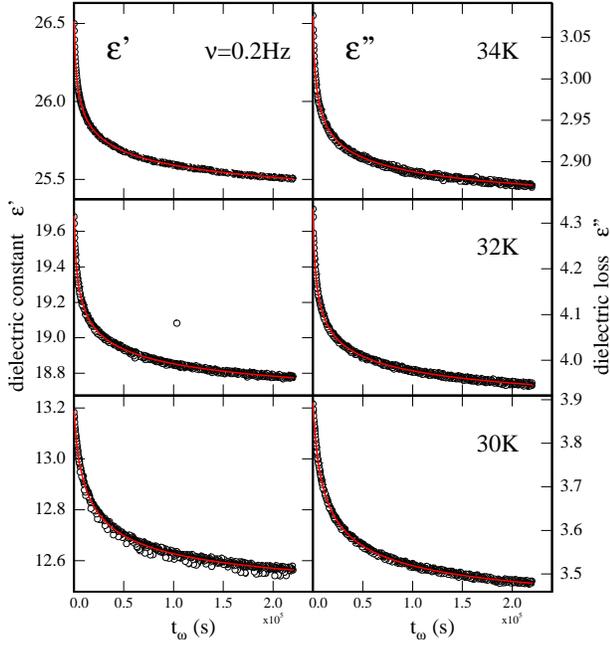}}}
\caption[]{(Color online) Isothermal time evolution of the
imaginary part of the 0.2 Hz-permittivity after quenching the
sample from 80 K to T$_a=30$ K, 32 K and 34 K, respectively. Solid
lines are fits after Eq.\@ (\ref{epsabfall}).}
\label{fitresults_final.eps}
\end{figure}

\begin{figure}
\centering
\resizebox{1.00\columnwidth}{!}{\rotatebox{0}{\includegraphics{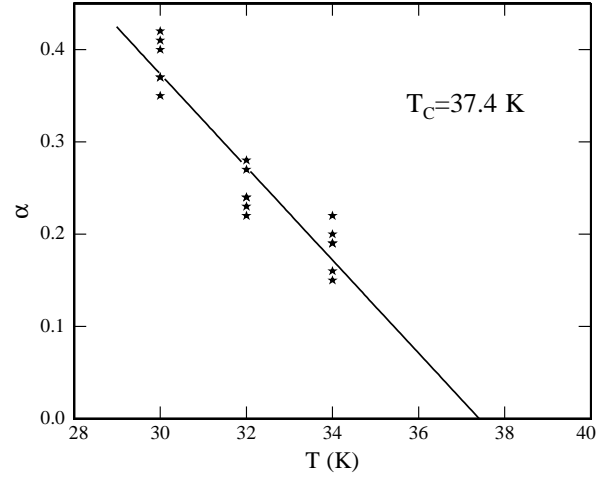}}}
\caption[]{Temperature dependence of the exponent $\alpha$ of the
power law for isothermal aging (Eq.\@ (\ref{epsabfall})). The
symbols ($\star$) indicate the values fitted for $\alpha'$ und
$\alpha''$ from Table \ref{tabelle} for the different frequencies
for the respective temperature. The linear fit points to a
critical temperature of T$_c$=37.4 K.} \label{tc.eps}
\end{figure}

\begin{figure}
\centering
\resizebox{0.75\columnwidth}{!}{\rotatebox{0}{\includegraphics{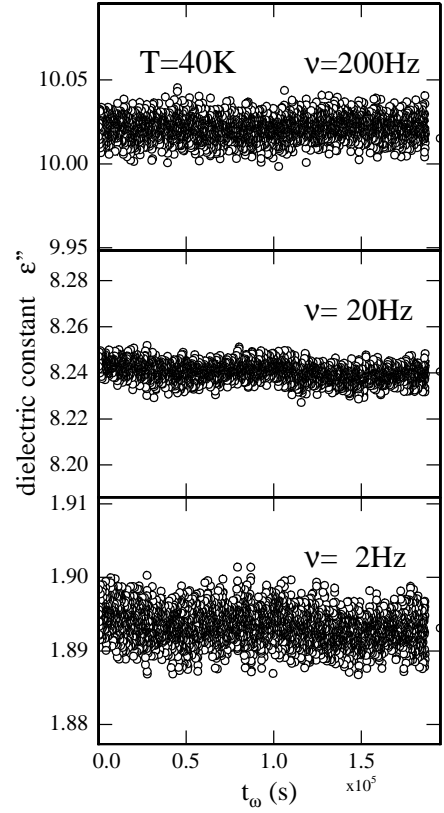}}}
\caption[]{Time evolution of the dielectric loss $\varepsilon''$
at T=40 K for $\nu$=2 Hz, 20 Hz and 200 Hz. $\varepsilon''$ stays
constant in time and there is no aging behavior.}
\label{aging_40k.eps}
\end{figure}

The first experiments refer to isothermal aging and serve to
characterizing the isothermal time evolution of the dielectric
constant. The system was heated to a high temperature T$=80$ K to
erase a possible thermal history. We then cooled down with a fast
rate of $|\Delta T/\Delta t|=7$ K/min to the measurement
temperature T$_a$. We chose T$_{a}=30$ K, 32 K and 34 K, where
$\varepsilon_c''(\nu)$ has reasonable high values for the selected
frequencies between $\nu$= 0.2 Hz and 200 Hz. The temperature was
then kept constant for the aging time $t_{a}$=3 days. During the
aging process, $\varepsilon_c'(\nu)$ and $\varepsilon_c''(\nu)$
were continuously measured for frequencies from $\nu=0.2$ Hz to
$\nu=200$ Hz. From their measurements on the dipolar glass
K$_{1-x}$Li$_x$TaO$_3$ Alberici {\em et
al.\@}\cite{Alberici:1997,Alberici:1997b,Alberici:1998,Alberici:2000}
find that the time evolution of the dielectric constant is well
described by a power law
\begin{equation}
\varepsilon'(\nu,t)=\varepsilon'(\nu,\infty)+\Delta\varepsilon'\left(\frac{t_w+t_0}{t_0}\right)^{-\alpha}.
\label{epsabfall}
\end{equation}
$\varepsilon'(\nu,\infty)$ is the asymptotic limit at infinite
time, $\Delta\varepsilon'$ measures the magnitude of the time
dependent part and $\alpha$ characterizes the decay. Because
cooling, heating and stabilizing the temperature always takes some
time, the start of an aging experiment is not well defined. The
time $t_0(>1)$ takes into account a possible delay of the
equilibrium of the temperature across the sample and the sample
cell. The evolution of both imaginary and real part of the
permittivity for all frequencies and all temperatures is well
described by Eq.\@ (\ref{epsabfall}). See Table \ref{tabelle} for
the fit parameters and Fig.\@ \ref{fitresults_final.eps} for the
$\nu$=0.2 Hz-data on the real and imaginary part of the
permittivity measured at T$_a$=30, 32, 34 K.
For each combination of T and $\nu$ we obtain about the same
values of $t_0$ and $\alpha$ from fits to $\varepsilon'(t)$ and
$\varepsilon''(t)$. On the other hand $t_0$ and $\alpha$ depend
strongly on T (and slightly on $\nu$). $t_0$ varies from about 900
s at 34 K to 5000 s at 30 K. The temperature equilibration time of
our setup is of the order of some minutes. $t_0$ is clearly larger
than that. Obviously the equilibration between the thermal bath
and the polar degrees of freedom of the sample takes somewhat
longer. The exponent $\alpha$ of the power law decay increases
with decreasing T. A fit of a linear T-dependence
\begin{equation} \alpha(T)=A|(T_c-T)/T|
\label{equalpha}\end{equation} (with a parameter A of A=-0.05.) to
the data (Fig.\@ \ref{tc.eps}) suggests a characteristic
temperature T$_c$ of about 38 K above which aging no longer
occurs.
\begin{table}
\begin{tabular}{ccccccccccc}
T (K) & $\nu$ (Hz) & $\varepsilon'_{\infty}$ &
$\Delta\varepsilon'$ & $\alpha'$ & t$_0'$ &
$\varepsilon''_{\infty}$ & $\Delta\varepsilon''$ & $\alpha''$ &
t$_0''$ 
\\\hline

30 & 0.2 & 12.37 & 0.81 & 0.37 & 4800 & 3.37 & 0.53 & 0.40 &
    4800 \\
   &  2  & 7.95  & 0.25 & 0.41 & 4750 & 2.28 & 0.24 & 0.42 &
    5250 \\
   & 20  & 5.67  & 0.05 & 0.35 & 5250 & 1.24 & 0.08 & 0.37 &
    5250 \\
32 & 0.2 & 18.40 & 1.28 & 0.24 & 1400 & 3.83 & 0.49 & 0.28 &
    1400 \\
   &  2  & 11.8  & 0.50 & 0.22 & 1600 & 3.97 & 0.39 & 0.27 &
    1600 \\
   & 20  & 7.22  & 0.12 & 0.24 & 1600 & 2.27 & 0.15 & 0.23 &
    1500 \\
34 & 0.2 & 24.96 & 1.56 & 0.19 &  900 & 2.79 & 0.29 & 0.22 &
    900 \\
   &  2  & 18.09  & 1.01 & 0.15 &  900 & 5.36 & 0.47 & 0.20 &
    900 \\
   & 20  & 10.31  & 0.26 & 0.19 &  900 & 4.01 & 0.30 & 0.16 &
    700 \\
\end{tabular}
\caption[]{Fitparameter for the frequencies $\nu$=0.2 Hz, 2Hz and
20 Hz for temperatures T=30 K, 32 K and 34 K. The $'$-values
indicate results for the real-, the $''$-values for the imaginary
part.
$\delta$=$\Delta\varepsilon''/(\varepsilon''_{\infty}+\Delta\varepsilon'')$
gives the relative strength of the aging process in the imaginary
part.}\label{tabelle}
\end{table}
Also experimentally we find a temperature of about T=38 K above
which we do not observe aging behavior. This is confirmed by
direct observation: Fig.\@ \ref{aging_40k.eps} shows an aging
experiment at T=40 K for frequencies $\nu$ of $\nu$=2 Hz, 20 Hz
and 200 Hz where the dielectric loss remains constant in time.
Because the effect of aging is more pronounced in the imaginary
part $\varepsilon_c''(\nu)$, we will in the following experiments
concentrate on this.

\subsection{Scaling experiments\label{scaling}}
\begin{figure}
\centering
\resizebox{1.00\columnwidth}{!}{\rotatebox{0}{\includegraphics{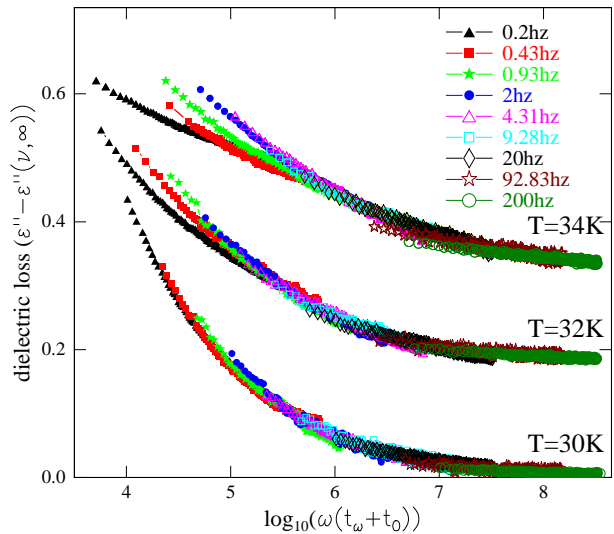}}}
\caption[]{(Color online)
$\varepsilon''-\varepsilon''(\nu,\infty)$ vs. $\omega
(t_{w}+t_{0})$ in a linear-logarithmic scale ($\omega
t_{w}$-scaling) for the temperatures T$_a=30$ K, 32 K und 34 K and
frequencies from $\nu$ von 0.2 Hz to $\nu=200$ Hz.}
\label{skaling.eps}
\end{figure}

\begin{figure}
\centering
\resizebox{1.00\columnwidth}{!}{\rotatebox{0}{\includegraphics{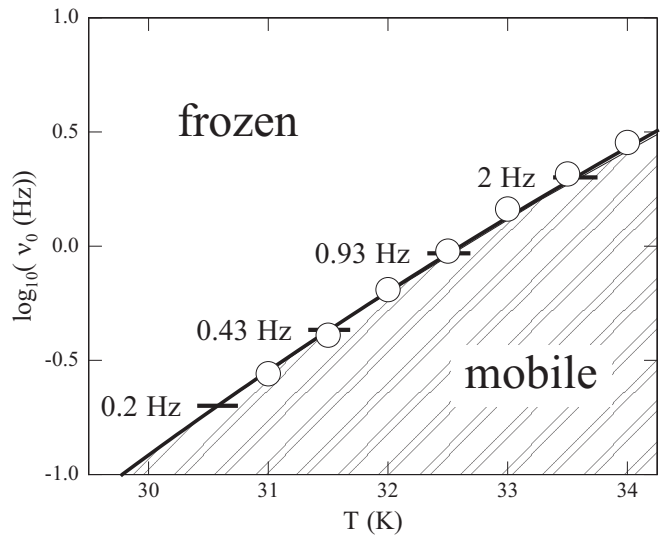}}}
\caption[]{Freezing line that separates frozen from mobile states.
The solid line is the Arrhenius-law after Fig. \@
\ref{paperbild.eps}, the points are corresponding experimental
data. Measurement frequencies are marked.}
\label{freezingline.eps}
\end{figure}

In spin-glasses an $\omega t_{w}$ scaling is observed, i.e.\@ the
aging curves for different frequencies fall on a single master
curve if the waiting time $t_w$ is multiplied by the applied
angular frequency $\omega$, ($\omega=2\pi\nu$). In orientational
glasses there often is a deviation from $\omega t_w$-scaling,
indicating that microscopic time scales are still relevant for the
aging process. In Fig.\@ \ref{skaling.eps},
$\varepsilon''-\varepsilon''(\nu,\infty)$ is plotted versus
$\omega (t_{w}+t_{0})$ in a linear-logarithmic scale. The values
for $\varepsilon''(\nu,\infty)$ and $t_0$ are the ones obtained
from the fits of Eq.\@ (\ref{epsabfall}) to the data.
For T$_a$=32 and 34 K the data taken at lower frequencies still do
show aging but no longer obey $\omega t_w$-scaling, the deviations
increase with decreasing $\nu$. The Arrhenius law of Fig.\@
\ref{paperbild.eps} translates into a boundary $\nu_0$(T) in the
T,$\nu$-plane separating the regimes in which the dipoles are
mobile or frozen in the sense that they can or cannot follow the
oscillations of the applied electric field (Fig.\@
\ref{freezingline.eps}). $\nu_0$ is the most probable relaxation
rate at the chosen temperature. Since the distribution of
relaxation rates extends over several decades\cite{Woll:2001} the
crossover from mobile to frozen states is rather gradual and we
still observe aging for $\nu$,T combinations well below the
$\nu_0$(T) boundary, that is deep in the mobile regime. It is only
at temperatures above about 38 K that aging effects can no longer
be resolved in the frequency range of the present study. See
Fig.\@ \ref{aging_40k.eps} for T=40 K. For the $\omega
t_w$-scaling, however, the distinction of frozen and mobile states
given by the $\nu_0$(T) boundary appears to be quite reliable. At
30 K all measuring frequencies chosen fall into the frozen regime,
$\nu>\nu_0$(T), thus the $\omega t_w$-scaling holds for all
measuring frequencies used. For 32 and 34 K only data with
$\nu>\nu_0$(T) obey scaling.

\subsection{Rejuvenation and memory\label{rejuvenation}}
\begin{figure}
\resizebox{1.00\columnwidth}{!}{\rotatebox{0}{\includegraphics{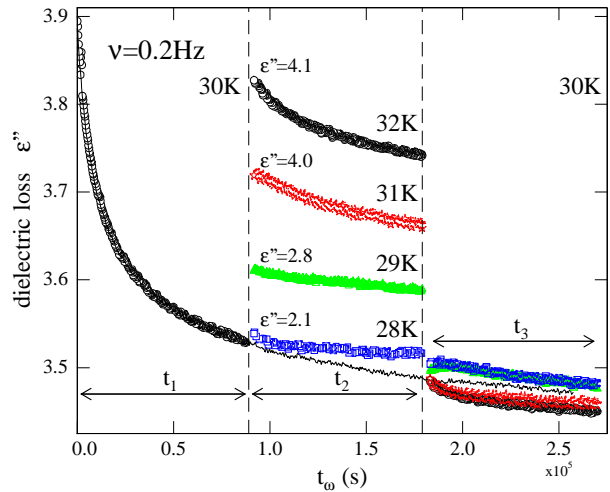}}}
\caption[]{(Color online) Temperature-cycling-experiment for
$\nu=0.2$ Hz and starting temperature T$=30$ K. The curves in
interval t$_2$ are shifted by arbitrary values of $\varepsilon''$,
the original values are given in the Figure.} \label{ts30_02.eps}
\end{figure}

If the system has aged at a temperature T$_1$ in the frozen regime
and the temperature is changed thereafter to a new value T$_2$, it
may forget that it has aged at T$_1$. This is rejuvenation. For
perfect rejuvenation the values of the quantity of interest (here
$\varepsilon''$) at T$_2$ are the same with and without aging at
T$_1$. Returning to T$_1$ the system may remember that it has aged
at this temperature before. In case of perfect memory the value of
$\varepsilon''$ after returning to T$_2$ is identical to the final
value of the previous aging at this temperature. If T$_2$ is well
in the mobile regime, rejuvenation and the loss of memory is
complete and on returning to T$_1$ the system is re-initialized.
Thus the effects of rejuvenation and memory are typically probed
by a temperature-time protocol consisting of isothermal aging over
time periods $t_1$, $t_2$, $t_3$ at temperatures T$_1$, T$_2$,
T$_1$ with T$_2$=T$_1$+dT. In the clathrate the $\nu$,T-regime in
which the dipole system is frozen but where on the other hand
$\varepsilon''$(T,$\nu$) is sufficiently large such that the
slight changes of $\varepsilon''$ brought forth by aging can be
studied with a reasonable resolution is rather narrow. We have
chosen $\nu$=0.2 Hz, T$_1$=30 K and dT$=+2,+1,-1,-2$ K in
combination with periods $t_1=t_2=t_3=30$ h (and realize that for
dT=2 K, T$_2$=32 K the system is already mobile if one strictly
refers to the $\nu_0$(T) boundary of Fig.\@
\ref{freezingline.eps}). The results are shown in Fig.\@
\ref{ts30_02.eps}, including a reference curve where aging has not
been interrupted by a T-excursion, that is with dT=0. The initial
state at the beginning of the first period has been approached by
rapid cooling out of the mobile regime at 80 K by cooling down
with a rate of 7 K/min. For the changes from T$_1$ to T$_2$ and
back to T$_1$ the rate was 1 K/min. The $t_2$-section of the
curves has been shifted vertically by arbitrary amounts in order
to present the results on a sufficiently expanded scale, but the
absolute values of $\varepsilon''$ in this interval can be
recovered since the start values are indicated in the Figs. The
results on the $t_1$-interval are of course independent on the
subsequent value of the temperature excursion dT and reproduce
what has been shown already in Fig.\@ \ref{fitresults_final.eps}.
The start values of the $t_2$-interval are close but not quite
identical to what is obtained by measuring the dielectric response
on continuous heating or cooling. $\varepsilon''$ at T=32 K for
$\nu$=0.2 Hz may serve as an example. Without previous aging a
value of about 4.3 is obtained as shown in Fig.\@
\ref{fitresults_final.eps}. After aging at 30 K for 30 h,
$\varepsilon''$ recovers from a value of about 3.5 at the end of
30 K-aging $t_1$-period to 4.1 at 32 K. Thus there is
rejuvenation, but rejuvenation is not complete meaning that the
age of the system at the temperature T$_2$ depends somewhat on the
former aging at the temperature T$_1$. As shown already in Ref.\@
[\onlinecite{Kityk:2002}] complete rejuvenation is only observed
for T-changes dT large enough. Bringing the sample back to
T$_1$=30 K in Fig.\@ \ref{ts30_02.eps} after the excursion to
T$_2$ the response is not re-initialized but the system remembers
that it has aged at this temperature before. In fact the aging
curves of the $t_3$-interval are quite close to the reference
curve for which aging has not been interrupted by a temporary
T-excursion. This statement holds independent of whether dT and
the change d$\varepsilon''$ of $\varepsilon''$ induced by changing
the temperature from T$_1$ to T$_2$ is positive or negative. A
closer inspection shows however that the curves of the
$t_3$-interval are slightly shifted with respect to the reference
curve, upward for dT$<$0, d$\varepsilon<$0 and downward for
dT$>$0, d$\varepsilon>$0. Note that the sign of shift is opposite
to that of d$\varepsilon$. Thus there is memory not only of the
$t_1$-period but also a residual one of the $t_2$-period.
Surprisingly it is not the sign of d$\varepsilon''$ what the
system remembers. We propose to refer instead to the age at the
temperature T$_1$ being modified by the excursion to the
temperature T$_2$. For this purpose the curves of the
$t_3$-interval are shifted horizontally until they coincide with
the reference curve. For dT$<$0 this shift is about $0.5\cdot
10^5$ s to the left. Thus intermediate aging at temperatures
somewhat lower than T$_1$ contributes to the age at T$_1$ but is
only about half as effective as the aging at T$_1$. For dT$>$0
aging at T$_2$ is more effective than aging at T$_1$ for the same
time span, the effective age is increased by t=$6.5\cdot 10^4$ s
for dT=1K and by $1.1\cdot 10^5$ for dT=2K.

\begin{figure}
\resizebox{1.00\columnwidth}{!}{\rotatebox{0}{\includegraphics{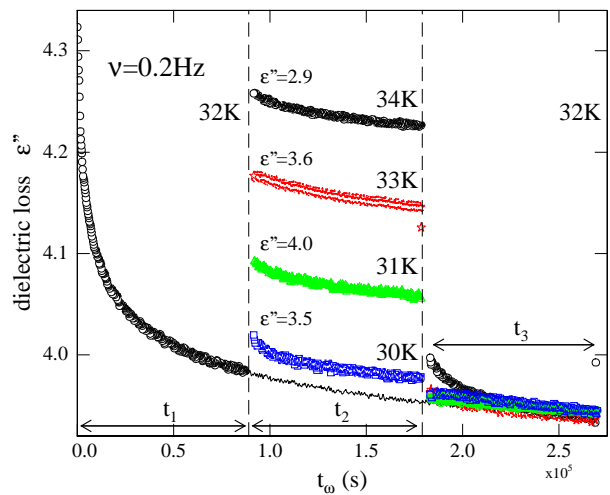}}}
\caption[]{(Color online) Analogous to Fig.\@ \ref{ts30_02.eps}
with a base temperature of T$_1$=32 K.} \label{ts32_02.eps}
\end{figure}

Figure \ref{ts32_02.eps} shows analogous data for a slightly
higher base temperature T$_1$ of 32 K. Most temperatures of
relevance for this experiment are already in the mobile regime of
Fig.\@ \ref{freezingline.eps}. Nevertheless the results are quite
similar to those shown in Fig.\@ \ref{ts30_02.eps}, except that
the starting value of the $t_3$-interval after an excursion to 34
K is higher than the final value of the $t_1$-interval. Obviously
an excursion to such a high temperature has partially erased the
memory and has lead to first traces of re-initialization.


\section{Discussion\label{Discussion}}
We find aging, rejuvenation and memory in
methanol-(x=0.73)-$\beta$-hydroquinone-clathrate similar to the
effects in spin-glasses. Although there is no experimental
evidence for a sharp glass {\it transition} in our system, we
observe a strong increase in the relaxation time with decreasing
temperature. Below 32K our system is out of equilibrium for all
relevant experimental time scales and the dynamics has pronounced
glassy features: In temperature cycle experiments we observed
rejuvenation as well as memory effects. Since our system is
randomly diluted the dipolar interactions between the moments lead
to the presence of a strong frustration due to a random mixture of
ferro- and antiferroelectric interactions.

Two different theoretical explanations for the aging phenomena
were put forward in the context of spin-glasses: In the droplet
theory \cite{Fisher:1988} rejuvenation is explained by the
assumption of a finite overlap length beyond which two equilibrium
configuration at different temperatures are decorrelated --- a
scenario that also goes under the name {\it chaos in
spin-glasses}\cite{Bray:1987,Fisher:1988}. Strong rejuvenation
effects necessitate in this simplified picture a very small
overlap length: If it would be much larger than the length-scales
that have been equilibrated during a particular waiting time, the
chaotic rearrangements due to the temperature shift become
invisible. However, there still might be some effect due to
dangerously irrelevant droplets whose free energy gaps are quite
small, as discussed in
\cite{YoshinoPRB:2002,Jonsson:2002,Jonsson:2003,Yoshino:2002}.

An alternative picture is the one presented in
\cite{Cugliandolo:1999,Bouchaud:2000,Berthier:2002}. Here
rejuvenation after a negative temperature shift comes from fast
modes equilibrated at $T_1$, but fall out of equilibrium and are
slow at $T_2$. Therefore, one should expect to see this phenomenon
if the equilibrated regions (on length scales $\le L_{T_1}(t_w)$)
are sufficiently different at the two temperatures. This mechanism
is obviously qualitatively different from the interpretation
involving the notion of temperature chaos (see above), which
implies that length scale smaller than the overlap length are
essentially unaffected by the temperature shift, while larger
length scales are completely reshuffled by the shift.

Moreover, within this picture the memory effect that we observe in
our temperature cycle experiments is a simple consequence of the
separation of time and length scales. When the system is at
$T_2<T_1$, rejuvenation involves very small length scales as
compared to the length scales involved in the aging at $T_1$. Thus
when the temperature is shifted back to $T_1$, the correlation of
length scale $L_{T_2}(t_s')$ nearly instantaneously re-equilibrate
at $T_1$. The memory is just stored in the intermediate length
scales, between $L_{T_2}(t_s')$ and $L_{T_1}(t_s)$.

The latter model accounts well for the aging effects observed in
spin-glasses. Strong random fields presumably destroy the spin-glass
character. In orientational glasses, in which the dipole dynamics is
usually strongly influenced by local crystal fields, which are
randomly distributed from site to site (''random fields''), aging
effects might be more appropriately described by a slow domain growth,
where the domain wall movement is impeded by the inhomogeneous crystal
fields. In this case the objects that are subject to
reconformations due to temperature changes are only the domain
walls whereas in spin glasses one expects that it is the whole bulk
of the domains.

In the present case local random fields arise from the partial
occupation. The neighborhood of statistically empty and filled
cavities leads to a random distortion of the crystal field in the
cages. Considering this special feature of dipolar glasses, aging
in the dipolar glass K$_{1-x}$Li$_x$TaO$_3$ has been explained by
Alberici {\em et al.\@}\cite{Alberici:1998,Alberici:2000} by a
domain model.  They assume the existence of polarization domains,
i.e.\@ regions of strongly coupled dipoles (even though the
interaction is random and the {\em domains} will not be comparable
to ordered domains as e.g.\@ in a ferromagnet). During the process
of domain growth the domain walls have to overcome energy barriers
via thermal activation, in close analogy what is supposed to
happen in random field systems \cite{Fisher:1986} as well as in
the droplet theory for spin glasses \cite{Fisher:1988}. Only
molecules in the domain walls contribute to the dielectric
response, molecules from inside the domains are already frozen.
The dielectric response $\varepsilon''$ is inversely proportional
to the domain size R, $\varepsilon''\propto 1/R(t)$, as the volume
fraction of wall molecules reduces when domains grow. For each
temperature T there is an equilibrium domain size $\xi(T)$. $\xi$
is the larger the lower the temperature, $\xi(T)\propto
1/\sqrt{T}$. All domains grow towards their equilibrium size; the
evolution stops when they have reached $\xi(T)$. When the
temperature is increased the equilibrium correlation length
decreases and it might happen that some of the domains have grown
beyond this lengths scale and have to adjust, i.e.\ shrink. For
domains with size $R(t,T)$ larger than $\xi(T)$ shrinking would be
a very slow process because they have to overcome large energy
barriers. A more efficient scenario would be the nucleation of
small domains (with sizes of the order of the lattice spacing)
within the larger domains and a rapid initial growth. In this way
nucleation would generate new domain walls giving rise to the
overshoot of the dielectric response in time interval $t_3$ of the
temperature cycling experiments in Fig.\@ \ref{ts30_02.eps}.

Matsuo and Suga \cite{Matsuo:1984} find for
$\beta$-hydroquinone-clathrates with various polar guest molecules
a proportionality between the transition temperature, where the
dipole system orders, and the square of the dipole moment, i.e.\@
between the thermal energy and the energy of the dipolar
interaction between the guest molecules. This suggests that the
dipole-dipole interaction between the guest molecules plays the
leading role for the dielectric behavior. From measurements of the
methanol(x=0.97), (x=0.84) and (x=0.79) samples\cite{Woll:2000},
the (hypothetical) transition temperature of the
methanol(x=0.73)-$\beta$-hydroquinone-clathrate can be estimated
to about T=38 K, where molecular reorientations should slow down
drastically. The critical temperature T$_c$ that arises from the
dielectric experiments agrees quite well with the experimental
finding suggesting that the collective dipolar interactions lead
to aging, memory and rejuvenation.

\section{Conclusions\label{Conclusions}}
We present dielectric measurements with different temperature-time
protocols to study aging effects in the dipolar glass
methanol\-(x=0.73)\--$\beta$\--hydroquinone\--clathrate. We find
aging, rejuvenation and memory effects, although there is no
experimental evidence for a sharp glass transition. The diluted
methanol clathrate is conceptually close to its magnetic pendants,
the spin-glasses. The electrical behavior is dominated by the
dipole-dipole interactions of the guest molecules but the coupling
of the pseudospins to random fields is not negligible. We argue
that the diluted methanol-clathrate takes an somehow intermediate
character between a true spin-glass and a pure random field
system. The influence of random fields can not be neglected as the
effects are much broader than the corresponding effects in spin
glasses (e.g.\@ in CdCr$_{1.7}$In$_{0.3}$S$_4$, see Ref.\@
[\onlinecite{Jonason:1998}]) and the overshoot in the time
interval t$_3$ is well explained by the domain model. Furthermore,
memory is preserved at moderate temperature increase opposed to
the observations in spin-glasses. On the other hand the $\omega
t_{w}$-scaling at low temperatures and the temperature dependence
of the fitting parameter $\alpha$ resemble the effects found in
spin-glasses and suggest a random bond dominated system. Aging,
rejuvenation and memory effects can just be observed below a
temperature at which the dipolar motions are expected to slow down
and collective dipolar interactions become dominant.

\bibliography{./aging}

\end{document}